\documentclass[12pt]{article}
\usepackage{amsmath,amsfonts,amssymb}

\begin{document}
\thispagestyle{empty}

\def\theequation{\arabic{section}.\arabic{equation}}
\def\a{\alpha}
\def\b{\beta}
\def\g{\gamma}
\def\d{\delta}
\def\dd{\rm d}
\def\e{\epsilon}
\def\ve{\varepsilon}
\def\z{\zeta}
\def\B{\mbox{\bf B}}

\newcommand{\h}{\hspace{0.5cm}}

\begin{titlepage}
\vspace*{1.cm}
\renewcommand{\thefootnote}{\fnsymbol{footnote}}
\begin{center}
{\Large \bf Leading finite-size effects on some three-point
correlators in $AdS_5\times S^5$}
\end{center}
\vskip 1.2cm \centerline{\bf Plamen Bozhilov} \vskip 0.6cm
\centerline{\sl Institute for Nuclear Research and Nuclear Energy}
\centerline{\sl Bulgarian Academy of Sciences} \centerline{\sl
1784 Sofia, Bulgaria}

\centerline{\tt plbozhilov@gmail.com}

\vskip 20mm

\baselineskip 18pt

\begin{center}
{\bf Abstract}
\end{center}
\h In the framework of the semiclassical approach, we find the
leading finite-size effects on the normalized structure constants
in some three-point correlation functions in $AdS_5\times S^5$,
expressed in terms of the conserved string angular momenta $J_1$,
$J_2$, and the worldsheet momentum $p_w$, identified with the
momentum $p$ of the magnon excitations in the dual spin-chain
arising in $\mathcal{N}=4$ SYM in four dimensions.

\end{titlepage}
\newpage

\def\nn{\nonumber}
\def\tr{{\rm tr}\,}
\def\p{\partial}
\newcommand{\bea}{\begin{eqnarray}}
\newcommand{\eea}{\end{eqnarray}}
\newcommand{\bde}{{\bf e}}
\renewcommand{\thefootnote}{\fnsymbol{footnote}}
\newcommand{\be}{\begin{equation}}
\newcommand{\ee}{\end{equation}}

\vskip 0cm

\renewcommand{\thefootnote}{\arabic{footnote}}
\setcounter{footnote}{0}


\setcounter{equation}{0}
\section{Introduction}
The correspondence between type IIB string theory on ${\rm
AdS}_5\times S^5$ target space and the ${\cal N}=4$ super
Yang-Mills theory (SYM) in four space-time dimensions, in the
planar limit, is the most studied example of the AdS/CFT duality
\cite{AdS/CFT}. A lot of impressive progresses have been made in
this field of research based on the integrability structures
discovered on both sides of the correspondence (for recent
overview on AdS/CFT integrability, see \cite{RO}).

Various classical string solutions play an important role in
testing and understanding the AdS/CFT correspondence. To establish
relations with the dual gauge theory, we have to take the
semiclassical limit of {\it large} conserved charges like string
energy $E$ and spins $S_{1,2}$ on ${\rm AdS}_5$ and angular
momenta $J_{1,2,3}$ on $S^5$\cite{GKP02}.

An example of such string solution is the so called "giant magnon"
, for which the energy $E$ and the angular momentum $J_1$ go to
infinity, but the difference $E-J_1$ is finite, while $S_{1,2}=0$,
$J_{2,3}=0$ \cite{HM06}. It lives on $R_t\times S^2$ subspace of
${\rm AdS}_5\times S^5$, and gave a strong support for the
conjectured all-loop $SU(2)$ spin chain, arising in the dual
$\mathcal{N}=4$ SYM, and made it possible to get a deep insight in
the AdS/CFT duality. This was extended to the giant magnon bound
state ($J_2\ne 0$), or {\it dyonic} giant magnon, corresponding to
a string moving on $R_t\times S^3$ and related to the complex
sine-Gordon model \cite{ND06}. Further extension to $R_t\times
S^5$ have been also worked out in \cite{KRT06}, where it was also
shown that such type of string solutions can be obtained by
reduction of the string dynamics to the Neumann-Rosochatius
integrable system. It can be used also for studding the {\it
finite-size effects}, related to the wrapping interactions in the
dual field theory \cite{Janikii}. From the string theory
viewpoint, the leading and even sub-leading finite size effect on
the giant magnon dispersion relation was first found and described
in \cite{AFZ06}. The case of leading finite-size effect on {\it
dyonic} giant magnon dispersion relation was considered in
\cite{HS0801}. There, the string theory result was compared with
the result coming from the $\mu$-term L\"uscher correction, based
on the $S$-matrix description. Both results coincide.

During the years, many important achievements concerning
correlation functions in the AdS/CFT context have been made.
Recently, interesting developments have been done by considering
general heavy string states \cite{Janik}-\cite{GLP1211}
\footnote{Some papers devoted to the field theory side of the
problem are also included here.}.

In \cite{PLB1107,PLB1108}, the three-point correlation functions
of finite-size (dyonic) giant magnons \cite{HM06,ND06} and three
different "light" states have been obtained. They are given in
terms of hypergeometric functions and several parameters. However,
it is important to know their dependence on the conserved string
charges $J_1$, $J_2$ and the worldsheet momentum $p$, because
namely these quantities are related to the corresponding operators
in the dual gauge theory, and the momentum of the magnon
excitations in the dual spin-chain. That is why, we are going to
find this dependence here. Unfortunately, this can not be done
exactly for the finite-size case due to the complicated dependence
between the above mentioned parameters and $J_1$, $J_2$, $p$.
Because of that, we will consider only the leading order
finite-size effects on the three-point correlators. In this paper,
we will restrict ourselves to the case of $AdS_5\times S^5$/
$\mathcal{N}=4$ SYM duality.

The paper is organized as follows. In Sec. 2, we first give a
short review of the giant magnon solution. Then, we explain the
limitations under which the three-point correlation functions
considered here are computed and give the exact results in the
semiclassical limit. Sec. 3 is devoted to the computation of the
leading order finite-size effects on the three-point correlators
given in Sec. 2 in terms of the conserved string angular momenta
and the worldsheet momentum $p$. In Sec. 4 we conclude with some
final remarks.

\setcounter{equation}{0}
\section{Finite-size giant magnons and \\ three-point correlators}

\subsection{Review of the giant magnon solutions}

The denote with $Y$, $X$ the coordinates in $AdS_5$ and $S^5$
parts of the background $AdS_5\times S^5$. \bea\nn
&&Y_1+iY_2=\sinh\rho\ \sin\eta\ e^{i\varphi_1},\\ \nn
&&Y_3+iY_4=\sinh\rho\ \cos\eta\ e^{i\varphi_2},\\ \nn
&&Y_5+iY_0=\cosh\rho\ e^{it}. \eea The coordinates $Y$ are related
to the Poincare coordinates by \bea \nn &&Y_m=\frac{x_m}{z},\\ \nn
&&Y_4=\frac{1}{2z}\left(x^mx_m+z^2-1\right), \\ \nn
&&Y_5=\frac{1}{2z}\left(x^mx_m+z^2+1\right), \eea where $x^m
x_m=-x_0^2+x_ix_i$, with $m=0,1,2,3$ and $i=1,2,3$. We
parameterize $S^5$ as in \cite{Hernandez2}.

Euclidean continuation of the time-like directions to $t_e = it$ ,
$Y_{0e} = iY_0$ , $x_{0e} = ix_0$, will allow the classical
trajectories to approach the $AdS_5$ boundary $z=0$ when $\tau_e
\rightarrow \pm\infty$, and to compute the corresponding
correlation functions.

The dyonic finite-size giant magnon solution, where
$(\tau,\sigma)$ are the world-sheet coordinates, can be written as
($t=\sqrt{W}\tau$,\ $i\tau=\tau_e$) \bea\nn
&&x_{0e}=\tanh(\sqrt{W}\tau_e),\h x_i=0,\h
z=\frac{1}{\cosh(\sqrt{W}\tau_e)},\eea
\bea\label{fsgm2}
&&\cos\theta=\sqrt{\chi_p}\
dn\left(\frac{\sqrt{1-u^2}}{1-v^2}\sqrt{\chi_p}(\sigma-v\tau)\Big\vert
1-\epsilon\right) ,
\\ \nn &&\phi_1= \frac{\tau-v\sigma}{1-v^2}+\frac{vW}{\sqrt{1-u^2}\sqrt{\chi_p}(1-\chi_p)}\times
\\ \nn &&
\Pi\left(-\frac{\chi_p}{1-\chi_p}\left(1-\epsilon\right),
am\left(\frac{\sqrt{1-u^2}}{1-v^2}\sqrt{\chi_p}(\sigma-v\tau)\right)\Big\vert
1-\epsilon\right) \\ \nn &&\phi_2=
u\frac{\tau-v\sigma}{1-v^2},\eea where $\theta$ is the angle on
which the metric on $S^3\subset S^5$ depends, while $\phi_{1,2}$
are the isometric angles on it. $dn\left(\alpha\vert
1-\epsilon\right)$ is one of the Jacobi elliptic functions,
$\Pi\left(\alpha, \beta\vert 1-\epsilon\right)$ is the incomplete
elliptic integral of third kind, and $am(x)$ is the Jacobi
amplitude.
Let us also mention that $\chi_p$, $\chi_m$ are related to $u$,
$v$, $W$ parameters according to \bea\nn
&&\chi_p+\chi_m=\frac{2-(1+v^2)W-u^2}{1
-u^2},\\
\label{3eqs} &&\chi_p \chi_m=\frac{1-(1+v^2)W+(v W)^2}{1
-u^2},\eea and
\bea\label{de}
\epsilon\equiv\frac{\chi_m}{\chi_p}.\eea

For the finite-size dyonic giant magnon string solution, the
explicit expressions for the conserved quantities and the
worldsheet momentum $p$ can be written as \cite{AB1105} \bea\nn
&&\mathcal{E} =\frac{2\sqrt{W}(1-v^2)}
{\sqrt{1-u^2}\sqrt{\chi_p}}\mathbf{K} \left(1-\epsilon\right), \\
\label{cqsGM} &&\mathcal{J}_1= \frac{2\sqrt{\chi_p}}
{\sqrt{1-u^2}}\left[ \frac{1-v^2W}{\chi_p}\mathbf{K}
\left(1-\epsilon\right)-\mathbf{E} \left(1-\epsilon\right)\right],
\\ \nn &&\mathcal{J}_2= \frac{2u\sqrt{\chi_p}}
{\sqrt{1-u^2}}\mathbf{E} \left(1-\epsilon\right)\\
\label{pw}&&p=\frac{2v} {\sqrt{1-u^2}\sqrt{\chi_p}}
\left[\frac{W}{1-\chi_p}\Pi\left(-\frac{\chi_p}{1-\chi_p}(1-\epsilon)\bigg\vert
1-\epsilon\right) -\mathbf{K} \left(1-\epsilon\right)\right], \eea
where\footnote{The relation between the string tension $T$ and the
t'Hooft coupling $\lambda$ in the dual $\mathcal{N}=4$ SYM is $T
R^2=\sqrt{\lambda}/2\pi$, where $R$ is the common radius of
$AdS_5$ and $S^5$ subspaces. Here $R$ is set to 1.} \bea\nn
\mathcal{E}=\frac{2\pi
E}{\sqrt{\lambda}},\qquad\mathcal{J}_{1,2}=\frac{2\pi
J_{1,2}}{\sqrt{\lambda}}\eea are the string energy and the two
angular momenta. $\mathbf{K} \left(1-\epsilon\right)$, $\mathbf{E}
\left(1-\epsilon\right)$, and \\
$\Pi\left(-\frac{\chi_p}{1-\chi_p}(1-\epsilon)\bigg\vert
1-\epsilon\right)$ are the complete elliptic integrals of first,
second and third kind. As explained in \cite{AFZ06}\footnote{See
also \cite{HS0801} for the dyonic case.}, (\ref{pw}) should be
identified with the momentum of the magnon excitations in the spin
chain arising in the dual $\mathcal{N}=4$ SYM theory.

The dyonic giant magnon dispersion relation, including the leading
finite-size correction, can be written as \bea\label{dfsdr}
\mathcal{E}-\mathcal{J}_{1}=\frac{\sqrt{\lambda}}{2\pi}\left[
\sqrt{\mathcal{J}_2^2+4\sin^2(p/2)} - \frac{\sin^4(p/2)}
{\sqrt{\mathcal{J}_2^2+4\sin^2(p/2)}}\ \epsilon\right],\eea where
\bea\label{epsilon} &&\epsilon=16
\exp\left[-\frac{2\left(\mathcal{J}_1 +
\sqrt{\mathcal{J}_2^2+4\sin^2(p/2)}\right)
\sqrt{\mathcal{J}_2^2+4\sin^2(p/2)}\sin^2(p/2)}{\mathcal{J}_2^2+4\sin^4(p/2)}
\right].\eea The second term in (\ref{dfsdr}) represents the
leading finite-size effect on the energy-charge relation, which
disappears for $\epsilon\to 0$, or equivalently
$\mathcal{J}_1\to\infty$. It is nonzero only for $\mathcal{J}_1$
finite.

The above two equalities are found under the following conditions
on the parameters \bea\nn 0<u<1,\h 0<v<1,\h 0<W<1,\h 0<\chi_{m}<
\chi_{p}<1.\eea

The case of finite-size giant magnons with one angular momentum
can be obtained by setting $u=0$, or $\mathcal{J}_2=0$, as can be
seen from (\ref{cqsGM}).


\subsection{Three-point correlation functions}

It is known that the correlation functions of any conformal field
theory can be determined  in principle in terms of the basic
conformal data $\{\Delta_i,C_{ijk}\}$, where $\Delta_i$ are the
conformal dimensions defined by the two-point correlation
functions
\begin{equation}\nn
\left\langle{\cal O}^{\dagger}_i(x_1){\cal O}_j(x_2)\right\rangle=
\frac{C_{12}\delta_{ij}}{|x_1-x_2|^{2\Delta_i}}
\end{equation}
and $C_{ijk}$ are the structure constants in the operator product
expansion
\begin{equation}\nn
\left\langle{\cal O}_i(x_1){\cal O}_j(x_2){\cal
O}_k(x_3)\right\rangle=
\frac{C_{ijk}}{|x_1-x_2|^{\Delta_1+\Delta_2-\Delta_3}
|x_1-x_3|^{\Delta_1+\Delta_3-\Delta_2}|x_2-x_3|^{\Delta_2+\Delta_3-\Delta_1}}.
\end{equation}
Therefore, the determination of the initial conformal data for a
given conformal field theory is the most important step in the
conformal bootstrap approach.

The three-point functions of two "heavy" operators and a "light"
operator can be approximated by a supergravity vertex operator
evaluated at the "heavy" classical string configuration
\cite{rt10,Hernandez2}: \bea \nn \langle
V_{H}(x_1)V_{H}(x_2)V_{L}(x_3)\rangle=V_L(x_3)_{\rm classical}.
\eea For $\vert x_1\vert=\vert x_2\vert=1$, $x_3=0$, the
correlation function reduces to \bea \nn \langle
V_{H}(x_1)V_{H}(x_2)V_{L}(0)\rangle=\frac{C_{123}}{\vert
x_1-x_2\vert^{2\Delta_{H}}}. \eea Then, the normalized structure
constants \bea \nn \mathcal{C}=\frac{C_{123}}{C_{12}} \eea can be
found from \bea \label{nsc} \mathcal{C}=c_{\Delta}V_L(0)_{\rm
classical}, \eea were $c_{\Delta}$ is the normalized constant of
the corresponding "light" vertex operator.

Recently, first results describing {\it finite-size} effects on
the three-point correlators appeared
\cite{AB1105,Lee:2011,AB11062,PLB1107,PLB1108}. This was done for
the cases when the "heavy" string states are {\it finite-size}
giant magnons, carrying one or two angular momenta, and for three
different choices of the "light" state:
\begin{enumerate}
\item{Primary scalar operators: $V_L=V^{pr}_j$} \item{Dilaton
operator: $V_L=V^d_j$} \item{Singlet scalar operators on higher
string levels: $V_L= V^q$}
\end{enumerate}

The corresponding (unintegrated) vertices are given by \cite{rt10}

\bea \label{prv} V^{pr}_j&=&\left(Y_4+Y_5\right)^{-\Delta_{pr}}
\left(X_1+iX_2\right)^j
\\ \nn
&&\left[z^{-2}\left(\p x_{m}\bar{\p}x^{m}-\p z\bar{\p}z\right) -\p
X_{k}\bar{\p}X_{k}\right],\eea where the scaling dimension is
$\Delta_{pr}=j$. The corresponding operator in the dual gauge
theory is $Tr\left( Z^j\right)$ \footnote{$Z$ is one of the three
complex scalars contained in $\mathcal{N}=4$ SYM.}. \bea
\label{dv} V^d_j&=&\left(Y_4+Y_5\right)^{-\Delta_d}
\left(X_1+iX_2\right)^j
\\ \nn
&&\left[z^{-2}\left(\p x_{m}\bar{\p}x^{m}+\p z\bar{\p}z\right) +\p
X_{k}\bar{\p}X_{k}\right], \eea where now the scaling dimension
$\Delta_d=4+j$ to the leading order in the large $\sqrt{\lambda}$
expansion. The corresponding operator in the dual gauge theory is
proportional to $Tr\left(F_{\mu\nu}^2 \ Z^j+\ldots\right)$, or for
$j=0$, just to the SYM Lagrangian. \bea\label{Vq}   V^q=
(Y_4+Y_5)^{- \Delta_q} (\p X_k \bar{\p} X_k)^q .\eea This operator
corresponds to a scalar {\it string} state at level $n=q-1$, and
to leading order in $\frac{1}{\sqrt{\lambda}}$ expansion
\bea\label{mc}
\Delta_q=2\left(\sqrt{(q-1)\sqrt{\lambda}+1-\frac{1}{2}q(q-1)}+1\right).\eea
The value $n=1 (q=2)$ corresponds to a massive string state on the
first exited level and the corresponding operator in the dual
gauge theory is an operator contained within the Konishi
multiplet. Higher values of $n$ label higher string levels.

The results obtained for the normalized structure constants
(\ref{nsc}), for the case of finite-size giant magnons in
$AdS_5\times S^5$, and the above three vertices, are as follows
\cite{PLB1107,PLB1108} \bea\label{c3prf}
&&\mathcal{C}_j^{pr}=\pi^{3/2}c_{j}^{pr}
\frac{\Gamma\left(\frac{j}{2}\right)}{\Gamma\left(\frac{3+j}{2}\right)}
\frac{\chi_p^{\frac{j-1}{2}}}{\sqrt{(1-u^2)W}}
\\ \nn &&\left[\left(1-W+j(1-v^2W)\right)
\
{}_2F_1\left(\frac{1}{2},\frac{1}{2}-\frac{j}{2};1;1-\epsilon\right)\right.
\\ \nn &&-\left.\left(1+j\right)\left(1-u^2\right)\chi_p
\
{}_2F_1\left(\frac{1}{2},-\frac{1}{2}-\frac{j}{2};1;1-\epsilon\right)\right],\eea

\bea\label{c3df} &&\mathcal{C}_j^d=2\pi^{3/2}c_{4+j}^{d}
\frac{\Gamma\left(\frac{4+j}{2}\right)}{\Gamma\left(\frac{5+j}{2}\right)}
\frac{\chi_p^{\frac{j-1}{2}}}{\sqrt{(1-u^2)W}}
\\ \nn &&\left[(1-u^2)\chi_p
\
{}_2F_1\left(\frac{1}{2},-\frac{1}{2}-\frac{j}{2};1;1-\epsilon\right)\right.
\\ \nn &&-\left.\left(1-W\right)
\
{}_2F_1\left(\frac{1}{2},\frac{1}{2}-\frac{j}{2};1;1-\epsilon\right)\right],\eea

\bea\label{cq} &&\mathcal{C}^q=c_{\Delta_q}\pi^{3/2}
\frac{\Gamma\left(\frac{\Delta_q}{2}\right)}{\Gamma\left(\frac{\Delta_q+1}{2}\right)}
\frac{(-1)^q\left[2-(1+v^2)W\right]^q}{(1-v^2)^{q-1}\sqrt{(1-u^2)W\chi_p}}
\\ \nn &&\sum_{k=0}^{q}\frac{q!}{k!(q-k)!}\left[-\frac{1-u^2}{1-\frac{1}{2}
(1+v^2)W}\right]^{k}\chi_p^{k}\
{}_2F_1\left(\frac{1}{2},\frac{1}{2}-k;1;1-\epsilon\right),\eea
where ${}_2F_1\left(a,b;c;z\right)$ is Gauss' hypergeometric
function.

\setcounter{equation}{0}
\section{Leading order finite-size effects}

As we already point out in the beginning, (\ref{cqsGM}),
(\ref{pw}), can not be solved {\it exactly} with respect to the
parameters involved, in order to express the relevant three-point
correlation functions in terms of the conserved charges and $p$.
That is why, we will consider here only the leading order
finite-size effects on the three-point correlators. This means
that we will consider the limit $\mathcal{J}_{1}$ large, i.e.
$J_1\gg \sqrt{\lambda}$, where the finite-size corrections to both
conformal dimensions and energies of string states have been
computed also from the L\"uscher corrections. Practically, the
problem reduces to consider the limit $\epsilon\to 0$, since
$\epsilon= 0$ corresponds to the infinite-size case, i.e.
$\mathcal{J}_{1}=\infty$. The relevant expansions of the
parameters are \cite{AB1105} \bea\nn
&&\chi_p=\chi_{p0}+\left(\chi_{p1}+\chi_{p2}\log(\epsilon)\right)\epsilon,
\h\chi_m=\chi_{m1}\epsilon,\h W=1+W_{1}\epsilon ,
\\
\label{Dpars} &&v=v_0+\left(v_1+v_2\log(\epsilon)\right)\epsilon,
\h u=u_0+\left(u_1+u_2\log(\epsilon)\right)\epsilon.\eea

The coefficients on the first line in (\ref{Dpars}) can be
obtained by using the equalities (\ref{3eqs}) and the definition
of $\epsilon$ (\ref{de}) to be
\bea\nn &&\chi_{p0}=1-\frac{v_0^2}{1-u_0^2}, \\
\nn &&\chi_{p1}=
\frac{v_0}{\left(1-v_0^2\right)\left(1-u_0^2\right)^2}
\Big\{v_0\left[(1-v_0^2)^2-3(1-v_0^2)u_0^2+2u_0^4-2(1-v_0^2)u_0u_1\right]
\\ \nn
&&-2\left(1-v_0^2\right)\left(1-u_0^2\right)v_1\Big\},\\
\label{chiW} &&\chi_{p2}=
-2v_0\frac{v_2+(v_0u_2-u_0v_2)u_0}{\left(1-u_0^2\right)^2}, \\
\nn &&\chi_{m1}=1-\frac{v_0^2}{1-u_0^2},
\\ \nn &&W_1=-\frac{(1-u_0^2-v_0^2)^2}
{(1-u_0^2)(1-v_0^2)}.\eea

The coefficients in the expansions of $v$ and $u$, we take from
\cite{PB10}, where for the case under consideration we have to set
$K_1=\chi_{n1}=0$, or equivalently $\Phi=0$. This leads to
\bea\label{zms}
&&v_0=\frac{\sin(p)}{\sqrt{\mathcal{J}_2^2+4\sin^2(p/2)}},\h
u_0=\frac{\mathcal{J}_2}{\sqrt{\mathcal{J}_2^2+4\sin^2(p/2)}}
\\ \nn &&v_1=\frac{v_0(1-v_0^2-u_0^2)}{4(1-u_0^2)(1-v_0^2)} \left[(1-v_0^2)(1-\log(16))
-u_0^2\left(5-v_0^2(1+\log(16))-\log(4096)\right)\right]
\\ \nn &&v_2=\frac{v_0(1-v_0^2-u_0^2)}{4(1-u_0^2)(1-v_0^2)} \left[1-v_0^2-u_0^2(3+v_0^2)\right]
\\ \nn &&u_1=\frac{u_0(1-v_0^2-u_0^2)}{4(1-v_0^2)} \left[1-\log(16)-v_0^2(1+\log(16))\right]
\\ \nn &&u_2= \frac{u_0(1-v_0^2-u_0^2)}{4(1-v_0^2)} (1+v_0^2).\eea

We need also the expression for $\epsilon$. It can be found from
the expansion of $\mathcal{J}_1$, and to the leading order is
given by (\ref{epsilon}).

\subsection{Giant magnons and primary scalar operators}

Let us first point out that (\ref{c3prf}) simplifies a lot when
$j$ is odd ($j=2 m+1, m=0,1,2,\ldots$). In that case, Gauss'
hypergeometric functions in (\ref{c3prf}) reduce to polynomials.
This results in \bea\label{cprodd} &&\mathcal{C}_{2
m+1}^{pr}=\pi^{3/2}c_{2 m+1}^{pr}
\frac{\Gamma\left(m+\frac{1}{2}\right)}{\Gamma\left(m+2\right)}
\frac{\epsilon^{m/2}\chi_p^m}{\sqrt{(1-u^2)W}}
\\ \nn &&\left[-2(m+1)(1-u^2)\sqrt{\epsilon}\ \chi_p
\ P_{m+1}\left(\frac{1+\epsilon}{2\sqrt{\epsilon}}\right)\right.
\\ \nn &&+\left.\left(1-W+(2m+1)(1-v^2 W)\right) \
P_{m}\left(\frac{1+\epsilon}{2\sqrt{\epsilon}}\right)\right],\eea
where $P_{n}(z)$ are Legendre's polynomials.

Since the corresponding operators in the dual gauge theory are of
the type $Tr\left(Z^j\right)$, we will restrict ourselves to
integer-valued $j$.

Let us start with the simpler case when $J_2=0$, or equivalently
$u=0$. Expanding (\ref{c3prf}) in $\epsilon$ and using
(\ref{Dpars}) - (\ref{zms}), one finds that\footnote{We use the
notation $\mathcal{C}_{j0}^{pr}$ in order to say that
$\mathcal{C}_{j}^{pr}$ are computed for the case $J_2=0$.}
\bea\label{Cpr0} &&\mathcal{C}_{10}^{pr}\approx 0,\h
\mathcal{C}_{20}^{pr}\approx
\frac{4}{3}c_{2}^{pr}\mathcal{J}_1\sin^2(p/2)\ \epsilon,
\\ \nn
&&\mathcal{C}_{j0}^{pr}\approx c_{j}^{pr}a_{j}\sin(p/2)^{j+1}\
\epsilon,\h j = 3, ..., 10,\eea where \bea\label{e0} \epsilon= 16
\exp[-2 - \mathcal{J}_1\csc(p/2)],\eea for the case under
consideration\footnote{This expression for $\epsilon$ comes from
(\ref{epsilon}) after setting $\mathcal{J}_2=0$.}. The numerical
coefficients $a_{j}$ are given by \bea\nn
a_{j}=\left(\frac{1}{4}\pi^{2},\frac{2^{4}}{3.
5},\frac{1}{16}\pi^{2},\frac{2^7}{3^2.5.7},\frac{3.5}{2^{9}}\pi^{2},
\frac{2^{10}}{3^3.5^2.7},\frac{5.7}{2^{11}}\pi^2,\frac{2^{14}}{3^2.5^2.7^2.11}\right).\eea

A few comments are in order. From (\ref{Cpr0}) one can conclude
that the $\mathcal{C}_{10}^{pr}$ and $\mathcal{C}_{20}^{pr}$ cases
are exceptional, while $\mathcal{C}_{j0}^{pr}$ have the same
structure for $j\ge 3$. $\mathcal{C}_{10}^{pr}\approx 0$ means
that the small $\epsilon$ - contribution to the three point
correlator is zero to the leading order in $\epsilon$.
$\mathcal{C}_{20}^{pr}$ is the only one normalized structure
constant of this type proportional to $\mathcal{J}_1$. It is still
exponentially suppressed by $\epsilon$. The common feature of
$\mathcal{C}_{j0}^{pr}$ in (\ref{Cpr0}) is that they all vanish in
the {\it infinite size} case, i.e., for $\epsilon =0$. This
property was established in \cite{Hernandez2}, and confirmed even
for the $\gamma$-deformed case in \cite{PLB1107}. Here, we
obtained the leading finite-size corrections to it.

Now, let us turn to the dyonic case, i.e. $J_2\ne 0$. Working in
the same way, but with $u\ne 0$, we derive

$j=1$:
\bea\label{c1pr}
&&\mathcal{C}_{1}^{pr}\approx
c_{1}^{pr}\frac{\pi^2}{16}\frac{\mathcal{J}_2^2 \csc(p/2)}{
[\mathcal{J}_2^2+4
\sin^2(p/2)]^{3/2}[\mathcal{J}_2^2+4\sin^4(p/2)]}\times
\\ \nn
&& \left\{8[\mathcal{J}_2^2+4
\sin^2(p/2)][\mathcal{J}_2^2+4\sin^4(p/2)]\right.
\\ \nn  &&+\left.\sin^2(p/2)\left[40+17\mathcal{J}_2^2+2\mathcal{J}_2^4
-20(3+\mathcal{J}_2^2)\cos(p)\right.\right. \\ \nn
&&+\left.\left.3(8+\mathcal{J}_2^2)\cos(2p)-4\cos(3p)-4\frac{\mathcal{J}_2^2+8
\sin^2(p/2)}{\mathcal{J}_2^2+4\sin^4(p/2)}\times\right.\right.
\\ \nn && \left.\left.\left(\mathcal{J}_1\sqrt{\mathcal{J}_2^2+4
\sin^2(p/2)}+\mathcal{J}_2^2+4
\sin^2(p/2)\right)\times\right.\right.
\\ \nn &&\left.\left.
\left(\mathcal{J}_2^2+4\sin^4(p/2)+2\sin^2(p)\right)\sin^2(p/2)\right]\epsilon\right\},\eea

$j=2$:
\bea\label{c2pr} &&\mathcal{C}_{2}^{pr}\approx
\frac{4}{3}c_{2}^{pr} \frac{1}{[\mathcal{J}_2^2+4
\sin^2(p/2)]^{3/2}[\mathcal{J}_2^2+4\sin^4(p/2)]}\times
\\ \nn
&&\left\{2\mathcal{J}_2^2[\mathcal{J}_2^2+4 \sin^2(p/2)]
\left[\mathcal{J}_2^2+4\sin^4(p/2)\right]-\sin^4(p/2)\times\right.
\\ \nn
&&\left.\left[20+3\mathcal{J}_2^2-2\mathcal{J}_2^4-2(15+2\mathcal{J}_2^2)\cos(p)
+(12+\mathcal{J}_2^2)\cos(2p)-2\cos(3p)\right.\right.
\\ \nn &&+
\left.\left.\frac{8}{\mathcal{J}_2^2+4\sin^4(p/2)}
\left(\mathcal{J}_1\sqrt{\mathcal{J}_2^2+4
\sin^2(p/2)}+\mathcal{J}_2^2+4
\sin^2(p/2)\right)\times\right.\right.
\\ \nn &&\left.\left.\left(-3+2(2+\mathcal{J}_2^2)\cos(p)
-\cos(2p)\right)\sin^4(p/2)\right]\epsilon\right\},\eea

$j=3$:
\bea\label{c3pr} &&\mathcal{C}_{3}^{pr}\approx c_{3}^{pr}
\frac{\pi^2}{256}\csc(p/2) \frac{[\mathcal{J}_2^2+4
\sin^2(p/2)]^{5/2}}{\mathcal{J}_2^2+4 \sin^4(p/2)}\times \\
\nn && \left\{48\mathcal{J}_2^2 \sin^2(p/2)
\frac{\mathcal{J}_2^2+4 \sin^4(p/2)}{[\mathcal{J}_2^2+4
\sin^2(p/2)]^{3}} -
\left[\frac{25\mathcal{J}_2^4}{[\mathcal{J}_2^2+4
\sin^2(p/2)]^{2}}\right.\right.\\ \nn
&&\left.\left.-\mathcal{J}_2^2 \frac{\mathcal{J}_2^2+4
\sin^4(p/2)}{[\mathcal{J}_2^2+4
\sin^2(p/2)]^{3}}\left(21-16\cos(p)-5\cos(2p)+8\mathcal{J}_2^2\right)
\right.\right.
\\ \nn &&\left.\left.
-\frac{3}{2}\mathcal{J}_2^6
\frac{11-12\cos(p)+\cos(2p)+6\mathcal{J}_2^2}{[\mathcal{J}_2^2+4
\sin^2(p/2)]^{4}} \right.\right. \\ \nn && \left.\left.
+\left(3\mathcal{J}_2^2 \left(\mathcal{J}_2^2+4
\sin^2(p/2)+\mathcal{J}_1\sqrt{\mathcal{J}_2^2+4
\sin^2(p/2)}\right)\times\right.\right.\right.
\\ \nn &&\left.\left.\left.\left(80+42\mathcal{J}_2^2+12\mathcal{J}_2^4
-\left(120+47\mathcal{J}_2^2-4\mathcal{J}_2^4\right)\cos(p)\right.\right.\right.\right.
\\ \nn &&\left.\left.\left.\left.+\left(8+\mathcal{J}_2^2\right)\left(6\cos(2p)-\cos(3p)\right)\right)\sin^4(p/2)
\right)\frac{1}{[\mathcal{J}_2^2+4
\sin^2(p/2)]^4[\mathcal{J}_2^2+4 \sin^4(p/2)]}\right.\right.
\\ \nn &&\left.\left.-\frac{20\mathcal{J}_2^4\sin^2(p)}{[\mathcal{J}_2^2+4
\sin^2(p/2)]^3}+\frac{3\mathcal{J}_2^4\sin^4(p)}{[\mathcal{J}_2^2+4
\sin^2(p/2)]^4}-8\left(\frac{\mathcal{J}_2^2+4
\sin^4(p/2)}{\mathcal{J}_2^2+4
\sin^2(p/2)}\right)^2\right]\epsilon\right\},\eea

$j=4$:
\bea\label{c4pr} &&\mathcal{C}_{4}^{pr}\approx
\frac{2}{45}c_{4}^{pr} \frac{[\mathcal{J}_2^2+4
\sin^2(p/2)]^{5/2}}{\mathcal{J}_2^2+4 \sin^4(p/2)} \\ \nn
&&\left\{\frac{32\mathcal{J}_2^2[\mathcal{J}_2^2+4
\sin^4(p/2)]\sin^2(p/2)}{[\mathcal{J}_2^2+4
\sin^2(p/2)]^3}-\left[\frac{17\mathcal{J}_2^4}{[\mathcal{J}_2^2+4
\sin^2(p/2)]^2}\right.\right. \\ \nn && \left.\left.
-\frac{1}{2}\mathcal{J}_2^2 \frac{\mathcal{J}_2^2+4
\sin^4(p/2)}{\mathcal{J}_2^2+4 \sin^2(p/2)]^3}
\left(39-32\cos(p)-7\cos(2p)+16\mathcal{J}_2^2\right)\right.\right.
\\ \nn &&\left.\left. -\mathcal{J}_2^6
\frac{11-12\cos(p)+\cos(2p)+6\mathcal{J}_2^2}{[\mathcal{J}_2^2+4
\sin^2(p/2)]^4}\right.\right. \\ \nn
&&\left.\left.+\left(2\mathcal{J}_2^2 \left(\mathcal{J}_2^2+4
\sin^2(p/2)+\mathcal{J}_1\sqrt{\mathcal{J}_2^2+4
\sin^2(p/2)}\right)\times\right.\right.\right.
\\ \nn &&\left.\left.\left.
\left(75+44\mathcal{J}_2^2+16\mathcal{J}_2^4
-2\left(58+23\mathcal{J}_2^2-4\mathcal{J}_2^4\right)\cos(p)\right.\right.\right.\right.
\\ \nn &&\left.\left.\left.\left.+ 4\left(13+\mathcal{J}_2^2\right)\cos(2p) -
2\left(6+\mathcal{J}_2^2\right)\cos(3p)+\cos(4p)\right)\sin^4(p/2)
\right)\times\right.\right.
\\ \nn &&\left.\left.\frac{1}{[\mathcal{J}_2^2+4
\sin^2(p/2)]^4[\mathcal{J}_2^2+4\sin^4(p/2)]}\right.\right.
\\ \nn &&\left.\left.-\frac{13\mathcal{J}_2^4\sin^2(p)}{[\mathcal{J}_2^2+4
\sin^2(p/2)]^3}+\frac{2\mathcal{J}_2^4\sin^4(p)}{[\mathcal{J}_2^2+4
\sin^2(p/2)]^4}-3\left(\frac{\mathcal{J}_2^2+4
\sin^4(p/2)}{\mathcal{J}_2^2+4
\sin^2(p/2)}\right)^2\right]\epsilon\right\}.\eea

In the four formulas above $\epsilon$ is given by (\ref{epsilon}).

\subsection{Giant magnons and dilaton operator}

The leading finite-size effect on the normalized structure
constant in the three-point correlator of two finite-size giant
magnon's states and zero-momentum dilaton operator ($j=0$), in the
limit $J_1\gg \sqrt{\lambda}$, has been considered in
\cite{AB1105}. Here, we will deal with the $j>0$ cases. Since the
corresponding operators in the dual gauge theory are proportional
to $Tr\left(F_{\mu\nu}^2 \ Z^j+\ldots\right)$, we will restrict
ourselves to integer-valued $j$.

When $j$ is odd ($j=2m+1, m=0,1,2,...$), the normalized structure
constants (\ref{c3df}) simplify to \bea\label{cdodd}
&&\mathcal{C}_{2 m+1}^d=2\pi^{3/2}c_{2 m+5}^{d}
\frac{\Gamma\left(m+\frac{5}{2}\right)}{\Gamma\left(m+3\right)}
\frac{\epsilon^{m/2}\chi_p^m}{\sqrt{(1-u^2)W}}
\\ \nn &&\left[(1-u^2)\sqrt{\epsilon}\ \chi_p
\
P_{m+1}\left(\frac{1+\epsilon}{2\sqrt{\epsilon}}\right)-\left(1-W\right)
\
P_{m}\left(\frac{1+\epsilon}{2\sqrt{\epsilon}}\right)\right].\eea

Expanding (\ref{c3df}) in $\epsilon$ and using (\ref{Dpars}) -
(\ref{zms}), one finds

$j=1$: \bea\nn
&&\mathcal{C}_1^d\approx\frac{3}{4}\pi^2 c_{5}^{d}\sin^3(p/2)
\Bigg\{\frac{1}{\sqrt{\mathcal{J}_2^2+4\sin^2(p/2)}}
\\ \nn
&&-\frac{1}{128\left(\mathcal{J}_2^2+4\sin^2(p/2)\right)^{3/2}
\left(\mathcal{J}_2^2+4\sin^4(p/2)\right)^{2}}\Big[\Big(840
+826\mathcal{J}_2^2+258\mathcal{J}_2^4-24\mathcal{J}_2^6 \\ \nn
&&-2\left(744+707\mathcal{J}_2^2+244\mathcal{J}_2^4+72\mathcal{J}_2^6\right)\cos(p)
\\ \nn &&+4\left(255+218\mathcal{J}_2^2+62\mathcal{J}_2^4-6\mathcal{J}_2^6\right)\cos(2p)
-\left(520+367\mathcal{J}_2^2+24\mathcal{J}_2^4\right)\cos(3p)
\\ \nn &&+2\left(92+47\mathcal{J}_2^2+3\mathcal{J}_2^4\right)\cos(4p)
-\left(40+11\mathcal{J}_2^2\right)\cos(5p)+4\cos(6p)\Big)
\\ \nn &&+8\mathcal{J}_1\sin^2(p/2)\sqrt{\mathcal{J}_2^2+4\sin^2(p/2)}\Big(
\left(8+19\mathcal{J}_2^2+12\mathcal{J}_2^4\right)\cos(p)+
\left(8-16\mathcal{J}_2^2\right)\cos(2p)
\\ \nn &&-\left(8+3\mathcal{J}_2^2\right)\cos(3p)
-2\left(5+5\mathcal{J}_2^2-2\mathcal{J}_2^4-\cos(4p)\right)\big)
\Big]\epsilon\Bigg\},\eea

$j=2$: \bea\nn &&\mathcal{C}_2^d\approx \frac{2^8}{3^2 5}
c_{6}^{d}\sin^4(p/2)
\Bigg\{\frac{1}{\sqrt{\mathcal{J}_2^2+4\sin^2(p/2)}}
\\ \nn
&&-\frac{1}{128\left(\mathcal{J}_2^2+4\sin^2(p/2)\right)^{3/2}
\left(\mathcal{J}_2^2+4\sin^4(p/2)\right)^{2}}\Big[\Big(210
+8\mathcal{J}_2^2\left(6-\mathcal{J}_2^2\right)\left(7+4\mathcal{J}_2^2\right)
\\ \nn
&&-8\left(63+84\mathcal{J}_2^2+38\mathcal{J}_2^4+16\mathcal{J}_2^6\right)\cos(p)
\\ \nn &&+\left(585+576\mathcal{J}_2^2+176\mathcal{J}_2^4-32\mathcal{J}_2^6\right)\cos(2p)
-4\left(115+84\mathcal{J}_2^2+4\mathcal{J}_2^4\right)\cos(3p)
\\ \nn &&+2\left(111+56\mathcal{J}_2^2+4\mathcal{J}_2^4\right)\cos(4p)
-4\left(15+4\mathcal{J}_2^2\right)\cos(5p)+7\cos(6p)\Big)
\\ \nn &&-8\mathcal{J}_1\sin^2(p/2)\sqrt{\mathcal{J}_2^2+4\sin^2(p/2)}\Big(
15+8\mathcal{J}_2^2-8\mathcal{J}_2^4-4\left(3+5\mathcal{J}_2^2
+4\mathcal{J}_2^4\right)\cos(p)
\\ \nn &&-
\left(12-8\mathcal{J}_2^2\right)\cos(2p)+4\left(3+\mathcal{J}_2^2\right)\cos(3p)
-3\cos(4p)\Big)\Big]\epsilon\Bigg\},\eea

$j=3$: \bea\nn &&\mathcal{C}_3^d\approx \frac{3. 5}{2^{5}}\pi^2
c_{7}^{d}\sin^5(p/2)
\Bigg\{\frac{1}{\sqrt{\mathcal{J}_2^2+4\sin^2(p/2)}}
\\ \nn
&&+\frac{1}{960\left(\mathcal{J}_2^2+4\sin^2(p/2)\right)^{3/2}
\left(\mathcal{J}_2^2+4\sin^4(p/2)\right)^{2}}
\Big[20\Big(256\left(13+15\cos(p)\right)\sin^{10}(p/2)
\\ \nn &&+288\mathcal{J}_2^2\left(5+7\cos(p)\right)\sin^{8}(p/2)
+\mathcal{J}_2^4\left(54+241\cos(p)+10\cos(2p)\right.
\\ \nn &&+15\left.\cos(3p)\right)\sin^{2}(p/2)+ 10\mathcal{J}_2^6\cos(p)
\left(5+3\cos(p)\right)\Big)
\\ \nn && +60\mathcal{J}_1\sin^2(p/2)\sqrt{\mathcal{J}_2^2+4\sin^2(p/2)}
\Big(20+6\mathcal{J}_2^2-12\mathcal{J}_2^4
\\ \nn
&&-\left(16+21\mathcal{J}_2^2+20\mathcal{J}_2^4\right)\cos(p)
-2\left(8-5\mathcal{J}_2^2\right)\cos(2p)
\\ \nn &&+\left(16+5\mathcal{J}_2^2\right)\cos(3p)
-4\cos(4p)\Big )\Big]\epsilon\Bigg\},\eea

$j=4$: \bea\nn &&\mathcal{C}_4^d\approx \frac{2^{11}}{3.5^2.7}
c_{8}^{d}\sin^6(p/2)
\Bigg\{\frac{1}{\sqrt{\mathcal{J}_2^2+4\sin^2(p/2)}}
\\ \nn
&&+\frac{1}{8192\left(\mathcal{J}_2^2+4\sin^2(p/2)\right)^{3/2}
\left(\mathcal{J}_2^2+4\sin^4(p/2)\right)^{2}} \Big[64\Big(294
+14\mathcal{J}_2^2-60\mathcal{J}_2^4+48\mathcal{J}_2^6 \\ \nn
&&-4\left(51-49\mathcal{J}_2^2-53\mathcal{J}_2^4-36\mathcal{J}_2^6\right)\cos(p)
\\ \nn &&-\left(435+8\mathcal{J}_2^2\left(61+19\mathcal{J}_2^2-6\mathcal{J}_2^4\right)\right)\cos(2p)
+2\left(305+209\mathcal{J}_2^2+6\mathcal{J}_2^4\right)\cos(3p)
\\ \nn &&-2\left(179+83\mathcal{J}_2^2+6\mathcal{J}_2^4\right)\cos(4p)
+2\left(53+13\mathcal{J}_2^2\right)\cos(5p)-13\cos(6p)\Big)
\\ \nn &&+512\mathcal{J}_1\sin^2(p/2)\sqrt{\mathcal{J}_2^2+4\sin^2(p/2)}\Big(
25+4\mathcal{J}_2^2-16\mathcal{J}_2^4-\left(20+22\mathcal{J}_2^2+24\mathcal{J}_2^4\right)\cos(p)
\\ \nn &&-4\left(5-3\mathcal{J}_2^2\right)\cos(2p)+2\left(10+3\mathcal{J}_2^2\right)\cos(3p)
-5\cos(4p)\Big)\Big]\epsilon\Bigg\}.\eea

In the four formulas above $\epsilon$ is given by (\ref{epsilon}).

Actually, we computed the normalized coefficients in the
three-point correlators up to $j=10$. However, since the
expressions for them are too complicated, we give here only the
results for the first two odd and two even values of $j$. Knowing
these expressions, the conclusion is that they have the same
structure for any $j$ in the small $\epsilon$ limit\footnote{The
only difference in that sense is that for $j$ odd an additional
overall factor of $\pi^2$ appears, as can be seen from the
formulas above.}. Namely

\bea\nn &&\mathcal{C}_j^d\approx A_j
c_{j+4}^{d}\sin^{j+2}\left(\frac{p}{2}\right)
\Bigg\{\frac{1}{\sqrt{\mathcal{J}_2^2+4\sin^2\left(\frac{p}{2}\right)}}
+\frac{a_j}{\left(\mathcal{J}_2^2+4\sin^2\left(\frac{p}{2}\right)\right)^{3/2}
\left(\mathcal{J}_2^2+4\sin^4\left(\frac{p}{2}\right)\right)^{2}}
\\ \label{av} &&
\Big[P_j^3(\mathcal{J}_2^2)+\mathcal{J}_1\sin^2\left(\frac{p}{2}\right)
\sqrt{\mathcal{J}_2^2+4\sin^2\left(\frac{p}{2}\right)} \
Q_j^2(\mathcal{J}_2^2)\Big]\epsilon\Bigg\}.\eea where $\epsilon$
is given in (\ref{epsilon}), $A_j$ and $a_j$ are numerical
coefficients, while $P_j^3(\mathcal{J}_2^2)$ and
$Q_j^2(\mathcal{J}_2^2)$ are polynomials of third and second order
respectively, with coefficients depending on $p$ in a
trigonometric way.

Now, let us restrict ourselves to the simpler case when
$\mathcal{J}_2=0$, i.e. giant magnon string states with one
(large) angular momentum $\mathcal{J}_1\neq 0$. Knowing the above
results for $1\le j \le 10$, one can conclude that the normalized
structure constants in the three-point correlators for any $j\ge
1$ in the small $\epsilon$ limit look
like\footnote{$\mathcal{C}_{j0}^d$ is used for $\mathcal{C}_{j}^d$
computed for $\mathcal{J}_2=0$ case.} \bea\label{av0}
&&\mathcal{C}_{j0}^d\approx \frac{A_j}{2} c_{j+4}^{d}
\sin^{j}\left(\frac{p}{2}\right)\left[\sin\left(\frac{p}{2}\right)\right.
\\ \nn
&&+\left.\left(B_{j0}\sin\left(\frac{p}{2}\right)+C_{j0}\sin\left(\frac{3p}{2}\right)
+D_{j0}\left(1+\cos(p)\right)\mathcal{J}_1\right)
e^{-2-\frac{\mathcal{J}_1}{\sin\frac{p}{2}}}\right],\eea where
\bea\nn &&B_{j0}=(-2^2,3,\frac{2.11}{3},11,\frac{2^3
3^2}{5},\frac{53}{3},\frac{2.73}{7},2^3 3,\ldots)\h \mbox{for}\h
j=(1,\ldots,8,\ldots),
\\ \nn &&C_{j0}=1+3j,\h D_{j0}=2(j+1).\eea

\subsection{Giant magnons and singlet scalar operators on higher string levels}
For that case, the expressions for the normalized structure
constants in the three-point correlation functions for {\it
dyonic} giant magnons are too long and complicated. That is why,
we will write down here the results for {\it finite-size} giant
magnon states only, i.e. for $\mathcal{J}_2=0$. Then, after small
$\epsilon$ expansion, one can find that (\ref{cq}) reduces
to\footnote{$\mathcal{C}^q_0\equiv \mathcal{C}^q$ computed for
$\mathcal{J}_2=0$.} \bea\label{vqe0} &&\mathcal{C}^q_0\approx
c_{\Delta_q}\frac{\sqrt{\pi}}{Aq_0}\frac{\Gamma\left(\frac{\Delta
q}{2}\right)}{\Gamma\left(\frac{1+\Delta q}{2}\right)}
\left\{Aq_1\sin(p/2)+Aq_2\mathcal{J}_1\right.
\\ \nn &&\left.+\left[\left(Aq_3+Aq_4\cos(p)\right)\sin(p/2)
+\left(Aq_5+Aq_6\cos(p)\right)\mathcal{J}_1\right.\right.
\\ \nn &&\left.\left.+Aq_7\csc(p/2)\left(1+\cos(p)\right)
\mathcal{J}_1^2\right]\epsilon\right\},\eea where $Aq_i$
$(i=0,1,...,7)$ are numerical coefficients, and for the case at
hand $\epsilon$ is given by (\ref{e0}).

This is the general structure of $\mathcal{C}^q_0$. The values of
$Aq_i$ we found are as follows $(q=1,\ldots,10)$
\bea\label{ncoeffs}
&&Aq_0=\left(8,24,60,420,2520,27720,180180,180180,3063060,116396280\right),
\\ \nn &&Aq_1=\left(16,-16,152,-632,7216,-55216,559304,-420312,10089896,-301915216\right),
\\ \nn &&Aq_2=\left(-8,24,-60,420,-2520,27720,-180180,180180,-3063060,116396280\right),
\\ \nn &&Aq_3=\left(2,-66,147,-2575,13446,-272694,1555993,-2484923,37469109,-2088496586\right),
\\ \nn &&Aq_4=\left(2,-10,171,-1027,15334,-144942,1747825,-1523631,41620821,-1396357874\right),
\\ \nn &&Aq_5=\left(-5,31,-\frac{187}{2},\frac{1837}{2},-6343,86653,-\frac{1256569}{2},
\frac{3}{2}490499,-\frac{27342361}{2},587890603\right),
\\ \nn &&Aq_6=\left(1,13,-\frac{97}{2},\frac{1207}{2},-4453,65863,-\frac{986299}{2},
\frac{3}{2}400409,-\frac{22747771}{2},500593393\right),
\\ \nn &&Aq_7=\left(-1,3,-\frac{15}{2},\frac{105}{2},-315,3465,-\frac{45045}{2},\frac{3}{2}15015,
-\frac{765765}{2},14549535\right).\eea

\setcounter{equation}{0}
\section{Concluding Remarks}
In this paper, in the framework of the semiclassical approach, we
computed the leading finite-size effects on the normalized
structure constants in some three-point correlation functions in
$AdS_5\times S^5$, expressed in terms of the conserved string
angular momenta $J_1$, $J_2$, and the worldsheet momentum $p_w$,
identified with the momentum $p$ of the magnon excitations in the
dual spin-chain arising in $\mathcal{N}=4$ SYM in four dimensions.
Namely, we found the leading finite-size effects on the structure
constants in three-point correlators of two "heavy" (dyonic) giant
magnon's string states and the following three "light" states:
\begin{enumerate}
\item{Primary scalar operators}; \item{Dilaton operator with
nonzero-momentum ($j\ge 1$)}; \item{Singlet scalar operators on
higher string levels}.
\end{enumerate}

A natural generalization of the above results would be to consider
the case of $\gamma$-deformed (or $TsT$-transformed) $AdS_5\times
S^5$ type IIB string theory background. Another possible issue to
investigate is the case of $AdS_4\times CP^3$ type IIA string
theory background, dual to $\mathcal{N}=6$ super
Chern-Simons-matter theory in three space-time dimensions (ABJM
model) and its $TsT$-deformations. We hope to report on these
soon.


\end{document}